\documentclass[aps,twocolumn,showpacs]{revtex4}
\usepackage{graphicx,amsmath}
\newcommand{\figwidth}{1.0\columnwidth}
\newcommand{\eq}[1]{Eq.(\ref{#1})}
\newcommand{\fig}[1]{Fig.~\ref{#1}}
\newcommand{\tab}[1]{Table~\ref{#1}}
\newcommand{\sect}[1]{Section~\ref{#1}}
\newcommand{\avg}[1]{{\langle#1\rangle}}
\newcommand{\olcite}[1]{Ref.~\onlinecite{#1}}
\newcommand{\ahum}[1]{``#1''}

\newcommand{\hx}{H_{\rm x}}
\newcommand{\eih}{\epsilon_\infty(H)}

\newcommand{\talpha}{{\tilde\alpha}}
\newcommand{\tbeta}{{\tilde\beta}}
\newcommand{\tgamma}{{\tilde\gamma}}

\begin{document}

\title{The isotropic-to-nematic transition in confined liquid crystals : an 
essentially non-universal phenomenon}

\author{J.M. Fish and R. L. C. Vink}

\affiliation{Institute of Theoretical Physics, 
Georg-August-Universit\"at G\"ottingen, Friedrich-Hund-Platz~1, 37077 
G\"ottingen, Germany}

\date{\today}

\begin{abstract} Computer simulations are presented of the isotropic-to-nematic 
transition in a liquid crystal confined between two parallel plates a distance 
$H$ apart. The plates are neutral and do not impose any anchoring on the 
particles. Depending on the shape of the pair potential acting between the 
particles, we find that the transition either changes from first-order to 
continuous at a critical film thickness $H=\hx$, or that the transition remains 
first-order irrespective of $H$. This demonstrates that the isotropic-to-nematic 
transition in confined geometry is not characterized by any universality class, 
but rather that its fate is determined by microscopic details. The resulting 
capillary phase diagrams can thus assume two topologies: one where the isotropic 
and nematic branches of the binodal meet at $H=\hx$, and one where they remain 
separated. For values of $H$ where the transition is strongly first-order the 
shift $\Delta \epsilon$ of the transition temperature is in excellent agreement 
with the Kelvin equation. Not only is the relation $\Delta \epsilon \propto 1/H$ 
recovered but also the prefactor of the shift is in quantitative agreement with 
the independently measured bulk latent heat and interfacial tension. 
\end{abstract}


\maketitle

\section{Introduction}

It is generally accepted that the first-order isotropic-to-nematic (IN) 
transition in liquid crystals confined between two parallel plates becomes 
continuous when the distance $H$ between the plates becomes small 
\cite{citeulike:4006298, citeulike:4067023, citeulike:3687059, 
citeulike:3687084, citeulike:4066932, citeulike:4057691}. Indeed, many 
simulations are consistent with this picture \cite{cleaver.allen:1993, 
lagomarsino.dogterom.ea:2003, dijkstra.roij.ea:2001, citeulike:4922416, 
citeulike:3683414} and show that the first-order IN transition terminates at a 
critical film thickness $\hx$. Some of these studies have also 
provided evidence of a continuous transition taking place when $H<\hx$. Note 
that as $H \to 0$ the system becomes effectively two-dimensional (2D). More 
recently, a (mathematically rigorous) proof appeared, showing that first-order 
IN transitions in 2D are also possible \cite{physrevlett.89.285702, 
enter.romano.ea:2006}. Inspired by this proof, computer simulations of liquid 
crystals in 2D were performed, which indeed uncovered strong first-order IN 
transitions too \cite{vink:2006*b, vink.wensink:2007}. Hence, the IN transition 
in confinement can be continuous, as well as first-order. Finally, there is the 
scenario of no transition occurring at all in thin films 
\cite{citeulike:5158181, citeulike:5158109}, not even a continuous transition of 
the Kosterlitz-Thouless (KT) type \cite{kosterlitz.thouless:1972}. Regarding 
experiments on confined liquid crystals, it has proved difficult to resolve 
continuous IN transitions in thin films \cite{citeulike:4067023, 
citeulike:4006159}. Pronounced coexistence between isotropic and nematic 
domains is typically observed \cite{citeulike:4006159, citeulike:2811025, 
citeulike:3991276}, which suggests that a transition does take place and that 
it is first-order.

The qualitatively different manifestations of the IN transition in confinement 
(continuous, first-order, absence) rule out any universality class for this 
transition. What remains of the IN transition in thin films is determined by 
microscopic detail. The only regime where some \ahum{agreement} may be obtained 
is in the bulk 3D limit $H \to \infty$. Here, one usually observes a first-order 
IN transition, with long-range order in the nematic phase. The transition thus 
breaks the rotational symmetry of the isotropic phase. At the mean-field level, 
this implies that the transition must be first-order \cite{citeulike:6170476}. 
We emphasize that fluctuations can change this result: even in 3D bulk, a 
genuine {\it continuous} IN transition is also possible \cite{citeulike:6170476, 
citeulike:6581608}. However, most bulk experiments yield a first-order IN 
transition, and so the mean-field approximation appears to be valid in 
this regime. As the film thickness $H$ decreases, fluctuations become 
increasingly important, and we expect three scenarios to unfold. In the first 
and most commonly accepted scenario, the IN transition becomes continuous when 
the film thickness drops below a critical thickness $\hx$. In addition, 
confinement is expected to destroy long-range order in the nematic phase, due to 
the Mermin-Wagner theorem \cite{physrevlett.17.1133}. Instead, quasi-long-range 
order may result, where the orientational correlations decay as a power law with 
distance. In the second (lesser known) scenario, the IN transition remains 
first-order irrespective of the film thickness, i.e.~all the way down to $H \to 
0$. In the third scenario, evidence for which was recently provided 
\cite{citeulike:5158181, citeulike:5158109}, the IN transition vanishes 
completely in the thin-film limit.

Given the three scenarios for the IN transition in confinement, all of which are 
qualitatively different, it is of fundamental interest to establish which 
\ahum{microscopic detail} is responsible for the scenario that ultimately 
occurs. The aim of this paper is to identify one possible mechanism, using 
computer simulations of a generalized Lebwohl-Lasher (LL) model. As it turns 
out, the generalized LL model is capable to reproduce all three scenarios, by 
tuning just a single parameter in the Hamiltonian. The effect of this parameter 
is to make the pair interaction \ahum{sharp and narrow}, meaning that particles 
interact when aligned but are otherwise rather indifferent to each other. 
Depending on this parameter, the crossover with decreasing film 
thickness from first-order to continuous behavior can be eliminated completely, 
and the IN transition remains first-order irrespective of $H$. 

The outline of this paper is as follows. We first introduce the generalized LL 
model and describe the simulation method. Next, we present new simulation data 
showing one example where the IN transition becomes continuous below a critical 
film thickness $\hx$, and a second example where the transition remains 
first-order irrespective of the film thickness. We do not consider the scenario 
where the transition vanishes below $\hx$ as this has recently been done 
elsewhere \cite{citeulike:5158181, citeulike:5158109}. A stringent test of the 
Kelvin equation, describing the shift of the transition temperature as a 
function of film thickness is also included. We end with a discussion 
and summary in \sect{the_end}.

\section{model and simulation method}
\label{model}

We consider a lattice model similar in spirit to the LL~model 
\cite{physreva.6.426}. To each site~$i$ of a 3D lattice, a 3D unit vector 
$\vec{d}_i$ (spin) is attached, which interacts with its nearest neighbors via
\begin{equation}\label{eqll}
 E = - \epsilon \sum_{\langle i,j \rangle} | \vec{d}_i \cdot \vec{d}_j |^p,
\end{equation}
with exponent $p$ and coupling constant $\epsilon$. In this work we absorb a 
factor of $1/k_B T$ in the coupling constant, with $k_B$ the Boltzmann constant 
and $T$ the temperature, and so $\epsilon$ plays the role of inverse 
temperature. The lattice is a $L \times L \times H$ rectangular box, with 
periodic boundary conditions in the lateral $L$ directions but not in the $H$ 
direction. The parameter $H$ thus plays the role of the film thickness; the 
minimum thickness that can be studied in this way equals $H=1$, corresponding to 
a single lattice layer. This setup is identical to the slab geometry used in 
earlier simulations of the confined LL~model \cite{cleaver.allen:1993}. Note 
that spins at the walls have a lower number of nearest neighbors, but that the 
walls are otherwise neutral, i.e.~we do not impose any anchoring conditions.

In the original LL~model the exponent of \eq{eqll} equals $p_{\rm LL}=2$. In the 
bulk limit $H \to \infty$, a (weak) first-order IN transition is observed 
\cite{citeulike:5091146, citeulike:3740162, physreva.6.426, citeulike:4197190, 
physrevlett.69.2803} at $\epsilon_\infty \approx 1.34$ \cite{note1}. In the 
thin-film limit $H=1$ recent simulations indicate the absence of any transition 
when $p=2$ \cite{citeulike:5158181, citeulike:5158109}. In this work we consider 
$p>2$. This modification is expected to enhance first-order phase transitions 
\cite{citeulike:5091592}, which may then even survive the limit $H \to 1$ 
\cite{vink:2006*b, vink.wensink:2007}. In line with previous work 
\cite{citeulike:5202797, physrevlett.69.2803, citeulike:3740162}, we analyze 
\eq{eqll} in terms of the histogram
\begin{equation}\label{eq:histo}
 P(E,S) \equiv P(E,S | H,L,\epsilon), 
\end{equation}
defined as the probability to observe a system with energy $E$ and nematic order 
parameter $S$ in a sample of thickness $H$, lateral extension $L$ and at inverse 
temperature $\epsilon$. The distribution is obtained by computer simulations 
using Wang-Landau \cite{wang.landau:2001, citeulike:278331} and transition 
matrix \cite{citeulike:202909} sampling; additional details pertaining to the 
present model are provided in \olcite{citeulike:5202797}. The nematic order 
parameter $S$ is defined in the usual way as the maximum eigenvalue of the 
orientational tensor
\begin{equation}
 Q_{\alpha \beta} = \frac{1}{2N} \sum_{i=1}^N
 \left( 3 d_{i\alpha} d_{\beta} - \delta_{\alpha\beta} \right),
\end{equation}
with $d_{i\alpha}$ the $\alpha$ component ($\alpha=x,y,z$) of the orientation 
$\vec{d}_i$ of the spin at site $i$, the sum over all $N=HL^2$ lattice sites 
and $\delta_{\alpha\beta}$ the Kronecker delta. In a perfectly aligned sample 
it holds that $S=1$, whereas an isotropic sample yields $S \to 0$ in the 
thermodynamic limit (hence, $S$ defined in this way is an intensive quantity). 
A final ingredient of this work is the use of finite-size scaling (FSS); needed 
because we seek thermodynamic limit properties. The thermodynamic limit of a 
film of thickness $H$ is defined by extending the lateral extension $L \to \infty$. 
In the bulk thermodynamic limit, both $H$ and $L$ are taken to infinity.

\section{Results}

Depending on the exponent $p$ in \eq{eqll}, we expect the first-order IN 
transition either to terminate at a critical film thickness $\hx$ or to remain 
first-order irrespective of $H$. The case $p=2$, i.e.~the original LL model, is 
an example of the former scenario. In the bulk limit one obtains a first-order 
transition \cite{citeulike:5091146, citeulike:3740162, physreva.6.426, 
citeulike:4197190, physrevlett.69.2803}, which terminates when the film 
thickness equals $\hx \sim 8 - 16$ lattice layers \cite{cleaver.allen:1993}. In 
the 2D limit $H=1$ no phase transition is observed for $p=2$ 
\cite{citeulike:5158181, citeulike:5158109}. We emphasize that the latter 
finding is not without some controversy, as previous other numerical studies of 
this system concluded that a phase transition does take place, namely a 
continuous transition of the KT type (see discussion in 
\olcite{citeulike:5158181}).

\subsection{crossover scenario}

\begin{figure}
\begin{center}
\includegraphics[width=\figwidth]{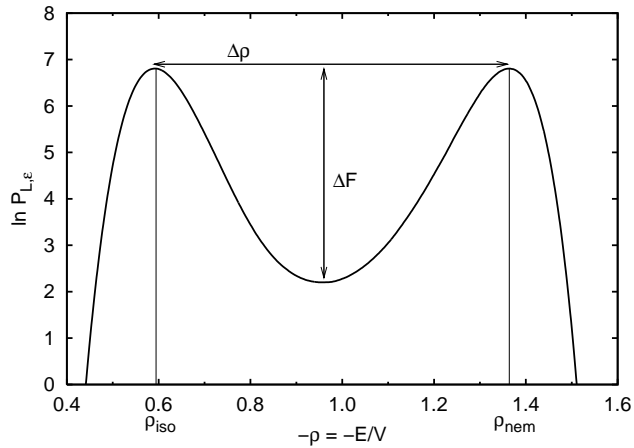}
\caption{\label{fig:op_dis} Logarithm of $P$ using $p=20$ in \eq{eqll} with 
$H=1$ and $L=25$. The value of $\epsilon$ has been chosen to give peaks of 
equal height. The free energy barrier, labeled $\Delta F$, is given as the 
difference between the peak maxima straddling the minimum. The distance 
labeled $\Delta \rho$ corresponds to the latent heat density. The distribution 
is plotted as a function of the negative energy density, such that the left peak 
corresponds to the isotropic phase and the right peak to the nematic phase.}
\end{center}
\end{figure}

\begin{figure}
\begin{center}
\includegraphics[width=\figwidth]{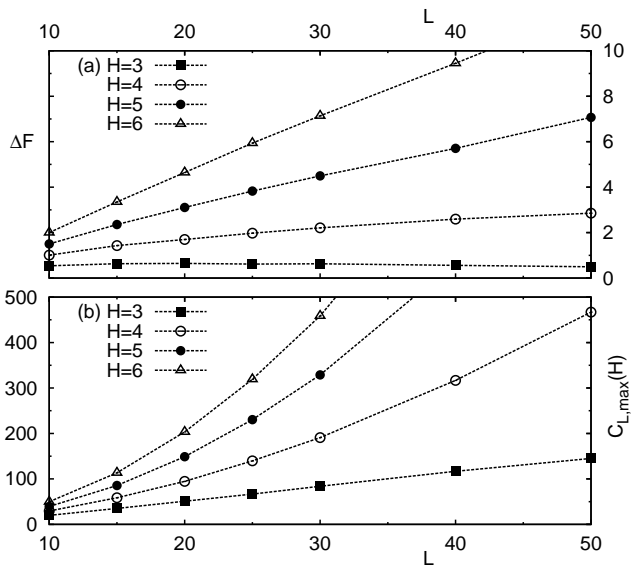}
\caption{\label{fig:xover} Evidence of the crossover scenario, whereby the IN 
transition ceases to be first-order below a critical film thickness; the results 
in this plot refer to $p=8$ in \eq{eqll}. The free-energy barrier $\Delta F$ 
versus the lateral film extension $L$ is plotted for several values of the film 
thickness $H$ in (a). For large $H$ the barriers increase linearly with $L$, 
consistent with a first-order transition; for smaller $H$ the barrier vanishes 
with increasing $L$. The maximum value of the specific heat 
versus $L$ is plotted in (b), again for several $H$.}
\end{center}
\end{figure}

We now consider \eq{eqll} using a larger exponent, $p=8$, to demonstrate that 
also a continuous IN transition is possible in thin films. To determine the 
order of the transition we use two FSS methods: the first was initially given 
by Lee and Kosterlitz \cite{physrevlett.65.137} and is based on the energy 
distribution $P(E)$, defined as the probability to observe a system with energy 
$E$
\[
 P(E) \equiv \int \int \delta(E-E') P(E',S') dE' dS',
\]
with $P(E,S)$ the joint distribution of \eq{eq:histo}.

At a first-order transition $P(E)$ becomes bimodal, see \fig{fig:op_dis} for 
an example, where the {\it logarithm} of the distribution is shown. For 
finite $L$ the bimodal structure persists over a range of $\epsilon$ values. 
As $L$ increases the range becomes smaller and in the thermodynamic limit 
$L \to \infty$ there is only one $\epsilon$ where $P(E)$ is bimodal, then 
featuring two $\delta$-peaks. Hence, for finite $L$ there is some freedom 
in choosing $\epsilon$ and in \fig{fig:op_dis} we have tuned $\epsilon$ 
such that the peaks are of equal height.

At a first-order transition the peak height, $\Delta F$ in $\ln P(E)$, see the 
vertical arrow in \fig{fig:op_dis}, corresponds to the free energy cost of 
interface formation \cite{binder:1982}. We therefore expect $\Delta F \propto 
L^{d-1}$, with $L$ the lateral extension of the film and $d=2$ (recall that 
films are effectively two-dimensional). To determine the order of the 
transition, Lee and Kosterlitz \cite{physrevlett.65.137} proposed to measure 
$\Delta F$ versus $L$, which should yield a {\it linear} increase for a film. 
Results for several values of $H$ are shown in \fig{fig:xover}(a). The data 
clearly indicate that the crossover scenario is taking place: for $H=6$ $\Delta 
F$ increases linearly with $L$, consistent with a first-order transition. In 
contrast, for $H=3$ $\Delta F$ vanishes for large $L$, implying the absence of a 
first-order transition.

To obtain the crossover thickness $\hx$ more accurately we use a second FSS 
method, based on the specific heat
\begin{equation}\label{eqcv}
 C = (\avg{E^2} - \avg{E}^2)/N,
\end{equation}
with $N=HL^2$ the number of lattice sites (volume). For given $L$ and $H$ a 
graph of $C$ versus $\epsilon$ reveals a maximum; the value of the maximum 
defines $C_{L, \rm max}(H)$. At a first-order transition the maximum scales with 
the volume of the system, that is $C_{L, \rm max}(H) \propto N$ 
\cite{citeulike:3610966}. In a film of fixed thickness $H$ this implies $C_{L, 
\rm max}(H) \propto L^\talpha$ with $\talpha_{\rm 1st}=2$. The result is shown 
in \fig{fig:xover}(b) for several values of the film thickness. For $H=6$ a fit 
yields $\tilde\alpha=2.00$, confirming that the transition is first-order. For 
$H=4$ we obtain $\tilde\alpha=1.74$, indicating that a first-order transition is 
absent. Hence, we conclude that the crossover thickness $\hx=5$. Precisely at 
$\hx$ a fit yields $\tilde\alpha=1.94$, which is still very close to the 
first-order value. Presumably for $H=5$ the IN transition is weakly first-order.

\begin{figure}
\begin{center}
\includegraphics[width=8cm]{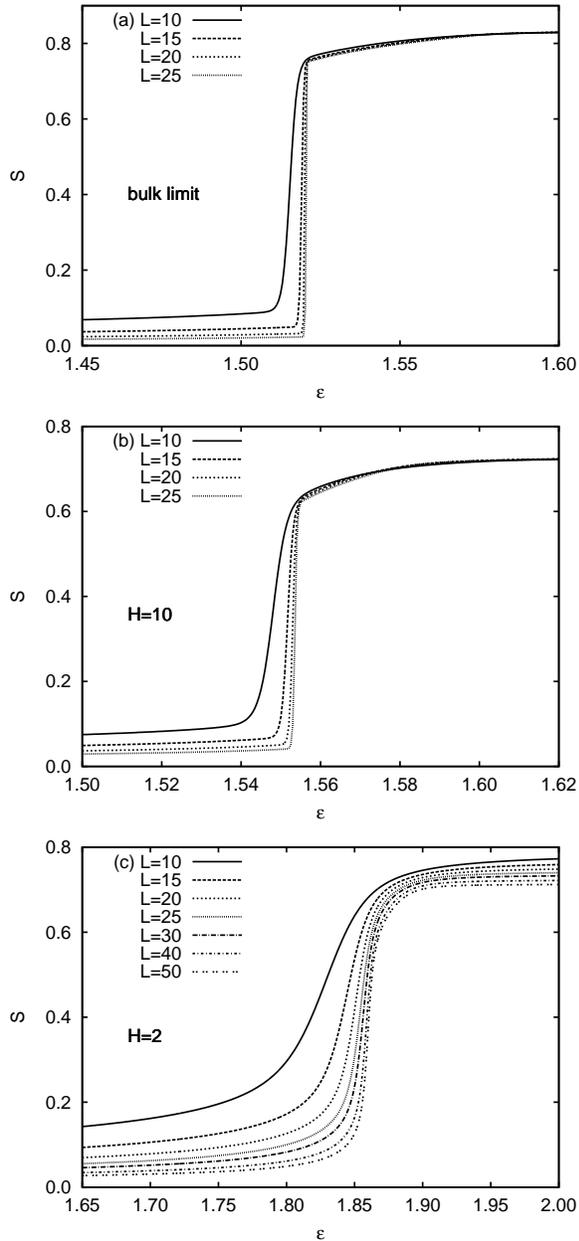}
\caption{\label{op} Variation of the nematic order parameter $S$ versus inverse 
temperature $\epsilon$ using $p=8$ in \eq{eqll} for several values of the film 
thickness $H$. In (a) we show the bulk result $H \to \infty$, whereas (b) and 
(c) were obtained in films of finite thickness $H$. Note that for (a) and (b) 
the IN transition is first-order while it has become continuous in (c).}
\end{center}
\end{figure}

For $H>\hx$, i.e.~where the transition is distinctly first-order, there is 
two-phase coexistence at the transition inverse temperature. It seems natural to 
characterize the phases with the nematic order parameter $S$. This approach is 
somewhat dangerous as confinement could destroy long-range nematic order in the 
thermodynamic limit: $\lim_{L \to \infty} S=0$ irrespective of $\epsilon$. For 
$H=1$ this follows rigorously from the Mermin-Wagner theorem 
\cite{physrevlett.17.1133}. The practical problem, affecting both simulations 
and experiments \cite{bramwell.holdsworth:1994}, is that the decay of $S$ with 
$L$ may be very slow. In fact, finite samples at low temperature typically 
reveal substantial order, even when the Mermin-Wagner theorem applies 
\cite{bramwell.holdsworth:1994}. The present simulations are no exception. Shown 
in \fig{op}(a) is $S$ versus $\epsilon$ in the bulk limit $H \to \infty$ for 
several system sizes $L$ (the bulk simulations were performed on a 3D cube of 
edge $L$ with periodic boundaries in all directions). A first-order IN 
transition taking place at $\epsilon \approx 1.52$ \cite{citeulike:5202797}, 
where $S$ jumps to a finite value, is clearly seen. More importantly, for 
$\epsilon$ above the transition, $S$ becomes independent of system size, at 
least on the scale of the graph; the latter is consistent with the formation of 
long-range nematic order, as expected in 3D. In \fig{op}(b) we show the 
corresponding result for a film of thickness $H=10$, which is still above the 
crossover thickness, and so the transition remains first-order. The behavior is 
similar to the bulk case, in the sense that $S$ \ahum{jumps} at the transition, 
and for large $\epsilon$ it appears to saturate at a finite value independent of 
the lateral film extension $L$. Hence, \fig{op}(b) provides no evidence of $S$ 
decaying to zero in the thermodynamic limit $L \to \infty$, but rather that the 
film supports long-range nematic order. If $S$ eventually does decay to zero, it 
is clear that huge system sizes, beyond the reach of any foreseeable simulation, 
are required to observe it.

\begin{figure}
\begin{center}
\includegraphics[width=\figwidth]{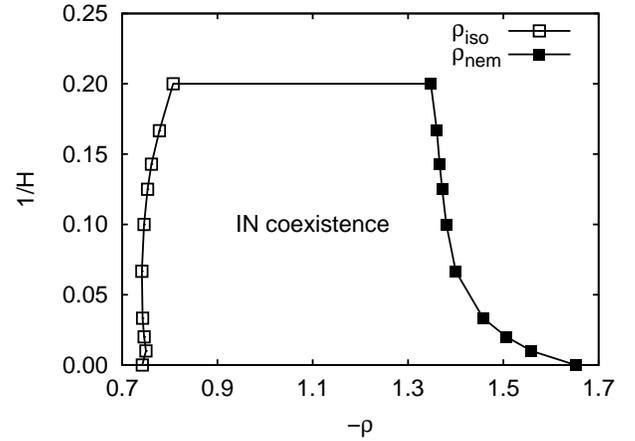}
\caption{\label{phase} Capillary phase diagram of \eq{eqll} using $p=8$. Shown 
is the variation of the coexisting phase densities $\rho_{\rm iso}(H)$ and 
$\rho_{\rm nem}(H)$ with the inverse film thickness $1/H$. The critical inverse 
thickness is at $1/\hx \sim 0.2$, above which the transition is no longer 
first-order and hence the two branches terminate. In the region between both 
branches coexistence between isotropic and nematic phases is observed.}
\end{center}
\end{figure}

To avoid these subtleties, we characterize the coexisting isotropic and nematic 
phases in the film with their energy densities $\rho_{\rm iso}(H)$ and 
$\rho_{\rm nem}(H)$ respectively. These are simply the peak positions 
in the energy distribution of \fig{fig:op_dis}. Recall that the latent heat of 
the transition equals ${\cal L}_L(H) = \rho_{\rm nem}(H) - \rho_{\rm iso}(H)$, 
where the subscript is a reminder of finite-size effects in the 
lateral film extension. The latent heat is related to the specific heat maximum 
\cite{citeulike:3610966}
\begin{equation}\label{eq:cvlh}
 {\cal L}_L(H) = \sqrt{4 C_{L, \rm max}(H) / N}
\end{equation}
and the extrapolation to $L \to \infty$ is performed assuming that ${\cal 
L}_\infty(H) - {\cal L}_L(H) \propto 1/N$. The average energy density
\[
 \rho_L(H) \equiv 
 \frac{\rho_{\rm iso}(H) + \rho_{\rm nem}(H)}{2} =
 \frac{1}{N} \int E P(E) \, d E
\]
obtained at the specific heat maximum is extrapolated analogously: 
$\rho_\infty(H) - \rho_L(H) \propto 1/N$. Once ${\cal L}_\infty(H)$ and 
$\rho_\infty(H)$ have been determined, the coexisting energy densities follow. 
The latter may then be plotted in a capillary phase diagram, see \fig{phase}, 
where the coexistence densities versus inverse film thickness $1/H$ are shown. 
Since the transition ceases to be first-order at the critical thickness $\hx$ 
the isotropic and nematic branches terminate.

\begin{figure}
\begin{center}
\includegraphics[width=8cm]{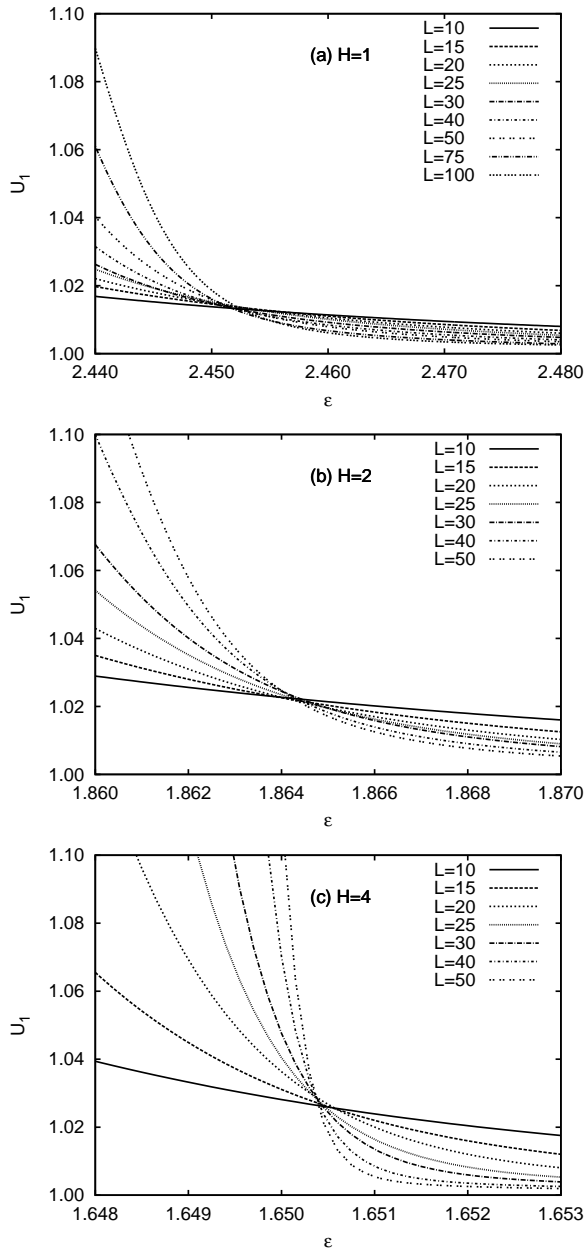}
\caption{\label{cumulant} Cumulant analysis of \eq{eqll} using $p=8$. Shown is 
$U_1$ versus $\epsilon$ using several values of the lateral film extension $L$, 
for (a) $H=1$, (b) $H=2$, and (c) $H=4$. The value of $\epsilon$ at the cumulant 
intersection yields the transition inverse temperature $\eih$ of 
the thermodynamic limit $L \to \infty$.}
\end{center}
\end{figure}

\begin{table}
\caption{\label{crit} Phase transition properties for $H<\hx$, for the 
continuous IN transition. Listed is the transition inverse temperature 
$\eih$, along with the exponents $\tilde\beta$ and $\tilde\gamma$, 
versus the film thickness $H$. The results refer to $p=8$ in \eq{eqll}.}
\begin{ruledtabular}
\begin{tabular}{cccc}
$H$ & $\eih$ & $\tilde\beta$ & $\tilde\gamma$ \\ \hline
1 & 2.450 & 0.19 & 1.63 \\
2 & 1.864 & 0.17 & 1.67 \\
3 & 1.716 & 0.15 & 1.71 \\
4 & 1.650 & 0.10 & 1.81 \\
\end{tabular}
\end{ruledtabular}
\end{table}

We now consider $H < \hx$. For $H=1$ and $p=2$ in \eq{eqll}, recent results 
\cite{citeulike:5158181, citeulike:5158109} indicate the absence of any phase 
transition (not even a continuous transition of the KT type). Part of the 
evidence is based on the failure of the Binder cumulant to intersect. At a 
continuous phase transition the ratio $U_1 = \avg{S^2} / \avg{S}^2$ becomes 
independent of system size \cite{binder:1981, citeulike:6170526}, where $S$ is 
the nematic order parameter. In simulations, this can be used to locate a 
continuous transition, by plotting $U_1$ versus $\epsilon$ for several system 
sizes $L$. At the transition inverse temperature $\eih$ of the 
film in the thermodynamic limit the curves for different lateral extensions $L$ 
are expected to intersect. While for $H=1$ and $p=2$ no intersections are found 
\cite{citeulike:5158109}, the result for $p=8$ is radically different, see 
\fig{cumulant}(a). Shown is $U_1$ versus $\epsilon$ using $H=1$ for several 
values of $L$. The curves clearly intersect and so we conclude that a 
continuous phase transition is taking place. This result strikingly illustrates 
the non-universality of the IN transition: whether a transition occurs for $H=1$ 
is determined by the exponent $p$ in \eq{eqll}, i.e.~a microscopic detail! Using 
$p=8$ we have verified that continuous transitions exist for all values of the 
film thickness $H < \hx$. The results for $H=2$ and $H=4$, where $H$ is 
approaching $\hx$, are shown in \fig{cumulant}(b) and~(c), both of which 
reveal cumulant intersections.

The fact that the cumulants intersect is a consequence of hyperscaling. At the 
transition inverse temperature $\eih$ the order parameter decays 
$\avg{S} \propto L^{-\tbeta}$, while the susceptibility $\chi = N ( \avg{S^2} - 
\avg{S}^2)$ diverges $\chi \propto L^\tgamma$. The exponents $\tbeta$ and 
$\tgamma$ are connected via the hyperscaling relation
\begin{equation}\label{eqhs}
 \tgamma + 2\tbeta = d, 
\end{equation} 
with spatial dimension $d=2$ for a film. This relation implies that the order 
parameter and its root-mean-square deviation scale $\propto L^x$ with the same 
exponent~$x$. Consequently, appropriately constructed cumulant ratios, such as 
$U_1$, become independent of $L$ whenever hyperscaling holds. By tuning the 
inverse temperature $\epsilon$ we have determined $\eih$ in our 
simulations by requiring that the scaling of $\avg{S}$ and $\chi$ with $L$ 
conforms to hyperscaling, i.e.~we numerically solved \eq{eqhs}. A solution to 
\eq{eqhs} for each $H<\hx$ could indeed be found; the resulting estimates of 
$\eih$, as well as the exponents $\tbeta$ and $\tgamma$, are 
listed in \tab{crit}. As expected, $\eih$ in \tab{crit} is close 
to the cumulant intersections of \fig{cumulant}, the discrepancy being less than 
0.1~\%. Note also that $\eih$ increases with decreasing $H$. The 
latter is consistent with the general tendency of confinement to lower phase 
transition temperatures. 

For $H=1$ the system has become 2D and the exponents reflect \ahum{pure} values, 
free from any crossover effects. Note that the exponents for $H=1$ deviate 
significantly from the XY~values $\tbeta_{XY}=1/8$ and $\tgamma_{XY}=7/4$ 
\cite{kosterlitz:1974}, strongly suggesting a different universality class. For 
$H>1$, the trend is that $\tbeta \to 0$, while $\tgamma \to 2$. Our 
interpretation is that, for $1<H<\hx$, one observes crossover scaling behavior 
\cite{citeulike:3687059}, governed by two competing fixed points: one being the 
first-order transition at $H=\hx$ and the other being the continuous transition 
at $H=1$. The exponents for $1<H<\hx$ are therefore \ahum{effective exponents}, 
with values between those of the $H=1$ system, and the \ahum{first-order} values 
$\tbeta_{\rm 1st} = 0$ and $\tgamma_{\rm 1st} = d = 2$ \cite{citeulike:3610966}. 
Note that effective exponents do not convey any fundamental information: if we 
were able to simulate arbitrarily large $L$ values arbitrarily close to the 
transition inverse temperature, the same exponents as for the $H=1$ system would 
be found.

\begin{figure}
\begin{center}
\includegraphics[width=\figwidth]{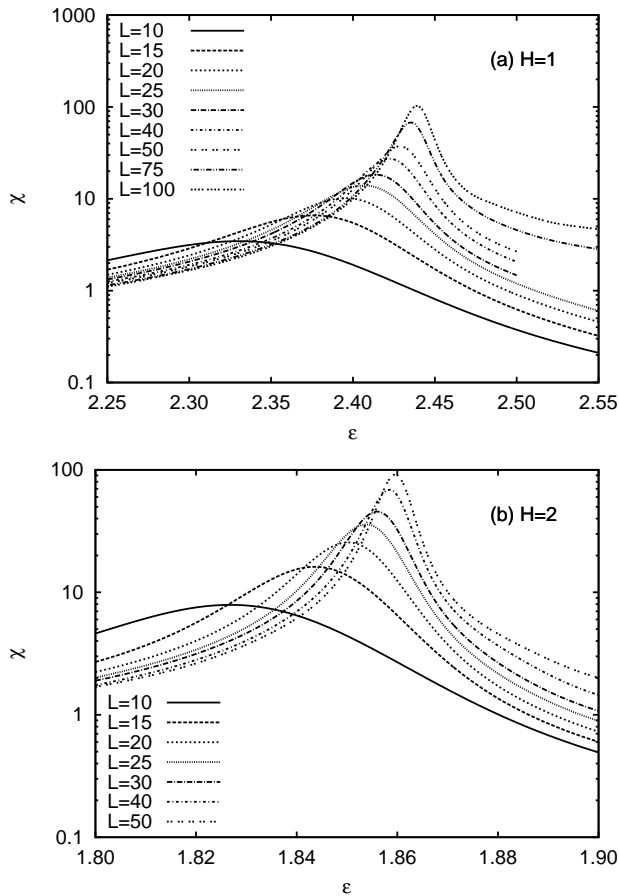}
\caption{\label{fig:sus} Variation of the susceptibility $\chi$ with inverse 
temperature $\epsilon$ for several values of the lateral film extension $L$, 
using film thicknesses (a) $H=1$ and (b) $H=2$. Both of these values $H<\hx$ and 
so the IN transition is continuous. Note the logarithmic vertical scale! The 
data were obtained using $p=8$ in \eq{eqll}.}
\end{center}
\end{figure}

\begin{figure}
\begin{center}
\includegraphics[width=\figwidth]{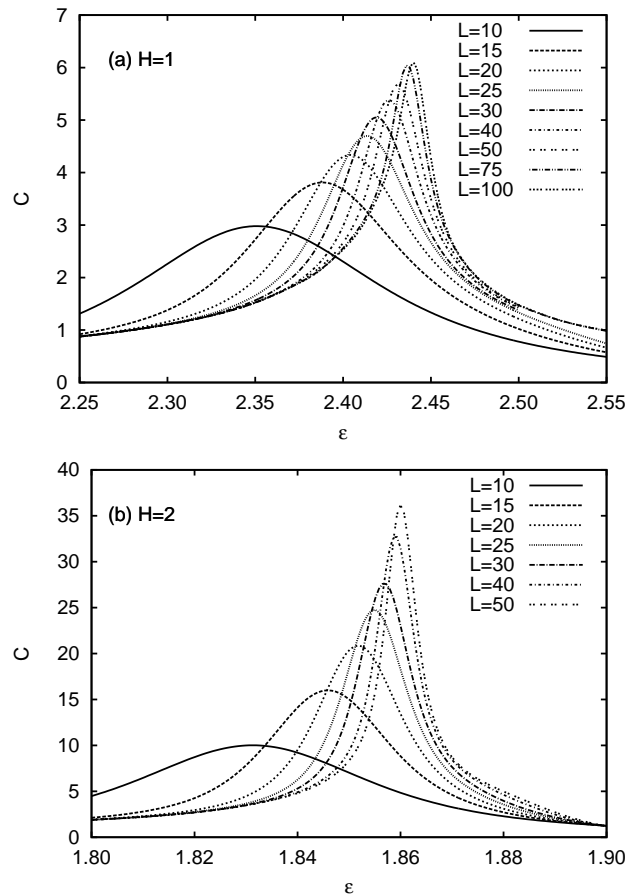}
\caption{\label{fig:cv} Variation of the specific heat $C$ with inverse 
temperature $\epsilon$ for several values of the lateral film extension $L$, 
using film thicknesses (a) $H=1$ and (b) $H=2$. Both of these values $H<\hx$ and 
so the IN transition is continuous. The data were obtained using $p=8$ in 
\eq{eqll}.}
\end{center}
\end{figure}

The important result to take from this analysis is that for $p=8$ in \eq{eqll} 
and $H<\hx$ a continuous phase transition is found; by enforcing hyperscaling 
the transition inverse temperature $\eih$ can be quite accurately 
obtained.

We now consider how the nematic order parameter $S$ depends on 
$\epsilon$ and $L$; a typical result is shown in \fig{op}(c) where $H=2$ was 
used. We note that $S$ increases with $\epsilon$ and that the slope 
$dS/d\epsilon$ reaches a maximum close to $\eih$. In contrast to 
the first-order transitions observed for $H>\hx$, $S$ does not saturate at high 
$\epsilon$ but decreases steadily with increasing $L$; this behavior is 
typical for all $H<\hx$. Our simulation data thus suggest the absence of 
long-range nematic order in the thermodynamic limit when $H<\hx$. This rules out 
a conventional critical point, since then the order parameter grows as a power 
law $S \propto t^\beta$, $t>0$, implying $S>0$ in the nematic phase, with 
distance from the transition
\begin{equation}\label{rel}
 t = \epsilon - \eih
\end{equation}
and $\beta$ the critical exponent of the order parameter. It is most likely, 
therefore, that the continuous transition we observe is a topological transition 
of the KT type \cite{kosterlitz.thouless:1972}. 

Consistent with the KT scenario is our previous result of the order parameter 
decaying $\avg{S} \propto L^{-\tbeta}$, and the susceptibility diverging $\chi 
\propto L^\tgamma$, whilst obeying hyperscaling. For completeness, we provide in 
\fig{fig:sus} some raw simulation data for the susceptibility. Clearly visible 
is that $\chi$ versus $\epsilon$ reveals a maximum, becoming more pronounced for 
increasing $L$. In principle, the inverse temperature $\epsilon_{L,\chi}(H)$ 
where the susceptibility reaches its maximum, in a film of thickness $H$ and 
lateral extension $L$, can be extrapolated using
\begin{equation}\label{fit}
 \epsilon_{L,\chi}(H) = \eih + \frac{b}{ \ln(L/c)^{1/\nu} },
\end{equation}
with non-universal constants $b$ and $c$, and where the exponent $\nu$ 
characterizes the exponential divergence of the correlation length $\xi \propto 
\exp(b t^\nu)$ for $t<0$, with $t$ given by \eq{rel}. For the XY~model it holds 
that $\nu_{XY}=1/2$, but since we did not recover XY~exponents in \tab{crit} 
the application of \eq{fit} requires that $\nu$ be fitted also, implying a 
4-parameter fit. We found that such a fitting procedure was numerically 
difficult to perform, and hence we did not determine $\eih$ in 
this manner. 

Finally, we note that also the specific heat, defined in \eq{eqcv}, is 
consistent with the KT scenario. Plotted in \fig{fig:cv} is the variation of $C$ 
with $\epsilon$ for several $L$, using two values of the film thickness. In both 
cases a maximum is revealed, but for $H=1$ it grows only weakly with $L$. This 
is consistent with a negative specific heat exponent, implying that $C$ remains 
finite in the thermodynamic limit, which agrees with the KT scenario. For $H=2$ 
we observe that $C$ already grows quite profoundly with $L$. We again attribute 
this to the crossover to a first-order transition where, ultimately, the 
specific heat maximum should scale $\propto L^\talpha$, with $\talpha_{\rm 
1st}=2$, see also \fig{fig:xover}(b).

\subsection{first-order transitions}

\begin{figure}
\begin{center}
\includegraphics[width=\figwidth]{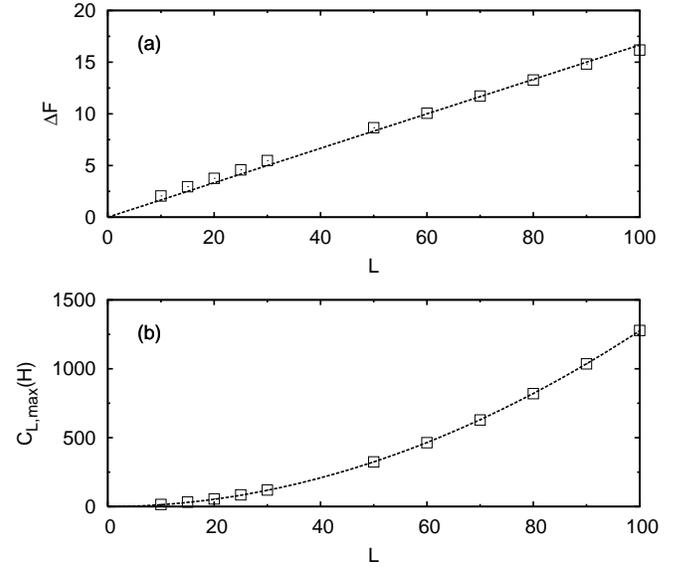}

\caption{\label{fig:first} Scaling analysis of \eq{eqll} using $p=20$ and film 
thickness $H=1$. In (a) we show the variation of the barrier $\Delta F$ versus 
$L$, while in (b) the specific heat maximum $C_{L, \rm max}(H)$ versus $L$ is 
shown. Both these results indicate a first-order phase transition, even though 
the system is purely 2D. The dashed line in (a) is the result of a linear fit 
through the origin. The curve in (b) is a fit to the form $C_{L, \rm max}(H) 
\propto L^\talpha$; we obtain $\talpha \approx 1.98$, which is very close to 
$\talpha_{\rm 1st}=2$ of a first-order phase transition in 2D.}

\end{center}
\end{figure}

\begin{figure}
\begin{center}
\includegraphics[width=\figwidth]{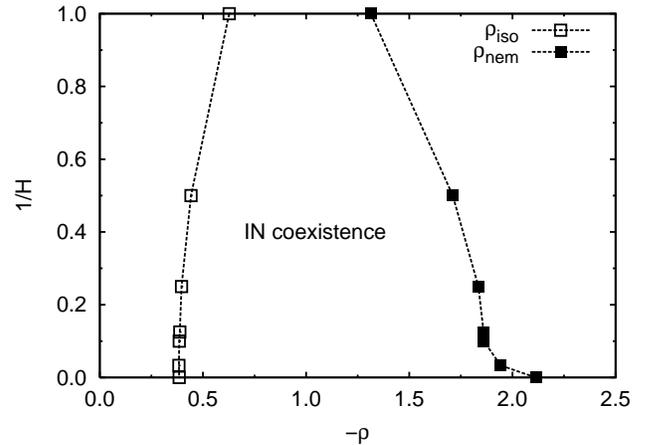}

\caption{\label{fig:p20_pd} Capillary phase diagram of \eq{eqll} using $p=20$. 
Shown is the variation of the coexisting energy densities $\rho_{\rm iso}(H)$ 
and $\rho_{\rm nem}(H)$ versus the inverse film thickness $1/H$. In this case no 
crossover occurs and the IN transition remains first-order irrespective of $H$. 
The isotropic and nematic branches of the binodal therefore do not terminate, 
but continue all the way to $H=1$.}

\end{center}
\end{figure}

We now consider the IN transition using $p=20$ in \eq{eqll}. In this case, the 
transition is strongly first-order, even in the thin-film limit. The application 
of the Lee-Kosterlitz scaling method for $H=1$ is shown in \fig{fig:first}(a), 
where the linear increase of the barrier $\Delta F$ with $L$ is clearly visible. 
The scaling of the specific heat maximum also confirms a first-order 
transition, see \fig{fig:first}(b), showing the expected quadratic 
dependence of $C_{L, \rm max}(H)$ on $L$. Since increasing the film thickness 
makes the transition more strongly first-order, it is clear that for $p=20$ no 
crossover can occur. In the capillary phase diagram, see \fig{fig:p20_pd}, the 
isotropic and nematic branches of the coexisting energy densities do not 
terminate, but continue all the way to $H \to 1$.

\subsection{Kelvin equation}

\begin{table}

\caption{\label{tab:first} Dependence of the transition inverse temperature 
$\eih$ on the film thickness $H$, for selected values of $H$ where 
the IN transition is first-order. Results are shown for exponents $p=8$ and 
$p=20$ in \eq{eqll}. The variation of $\eih$ with $H$ should 
follow the Kelvin equation, see \eq{eq:kelvin}. The bottom three lines list the 
bulk $(H \to \infty)$ transition inverse temperature $\epsilon_\infty$ the bulk 
latent heat density $\cal L_\infty$, and the bulk interfacial tension 
$\gamma_\infty$, which are required in order to compare to the Kelvin 
equation.}

\begin{ruledtabular}
\begin{tabular}{ccc|ccc}
$p=8$ & $H$ & $\eih$ & $p=20$ & $H$ & $\eih$ \\ \hline
& 5        & 1.614 & & 1 & 2.769 \\
& 6        & 1.593 & & 2 & 2.175 \\
& 7        & 1.578 & & 4 & 1.962 \\
& 8        & 1.568 & & 8 & 1.874 \\
& 10       & 1.555 & & 10 & 1.858 \\
& 15       & 1.540 & & 30 & 1.821 \\
& 30       & 1.528 & & & \\
& 50       & 1.525 & & &  \\
& 100      & 1.522 & & &  \\ \hline
& $\epsilon_\infty$ & 1.521 & & $\epsilon_\infty$ & 1.806 \\ 
& $\cal L_\infty$ & 0.909 & & $\cal L_\infty$ & 1.727 \\ 
& $\gamma_\infty$ & 0.06  & & $\gamma_\infty$ & 0.30 \\ 
\end{tabular}
\end{ruledtabular}
\end{table}

Finally, we study the variation of the inverse transition temperature $\eih$ 
with the film thickness for those cases where the IN transition is first-order. 
We expect $\eih$ to fit to the Kelvin equation \cite{citeulike:3687084} as
\begin{equation}\label{eq:kelvin}
 \Delta \epsilon \equiv 1 - \epsilon_\infty / \eih = 
 \frac{ 2 \gamma_\infty }{ {\cal L}_\infty H},
\end{equation}
where $\gamma_\infty$ is the bulk $(H \to \infty)$ interfacial tension, 
$\epsilon_\infty$ the bulk IN transition inverse temperature and $\cal 
L_\infty$ the bulk latent heat density. In deriving this equation complete 
wetting is assumed \cite{citeulike:3687084}. All quantities that appear in 
\eq{eq:kelvin} can, in principle, be extracted from finite-size simulation data 
with relative ease. For example, $\eih$ at a first-order transition can be 
obtained from $\epsilon_{L,k}(H)$; the latter is defined as the inverse 
temperature where the ratio of the peak areas in the energy distribution $P(E)$ 
equals $k$. For an optimal value $k = k_{\rm opt}$, which can be found using 
trial-and-error, the $L$-dependence in $\epsilon_{L,k}(H)$ becomes negligible 
and $\eih$ can be accurately obtained \cite{citeulike:5202797}. The resulting 
estimates of the transition inverse temperatures, for both $p=8$ and $p=20$, are 
provided in \tab{tab:first}, using only values of the film 
thickness where the transition is first-order.

\begin{figure}
\begin{center}
\includegraphics[width=\figwidth]{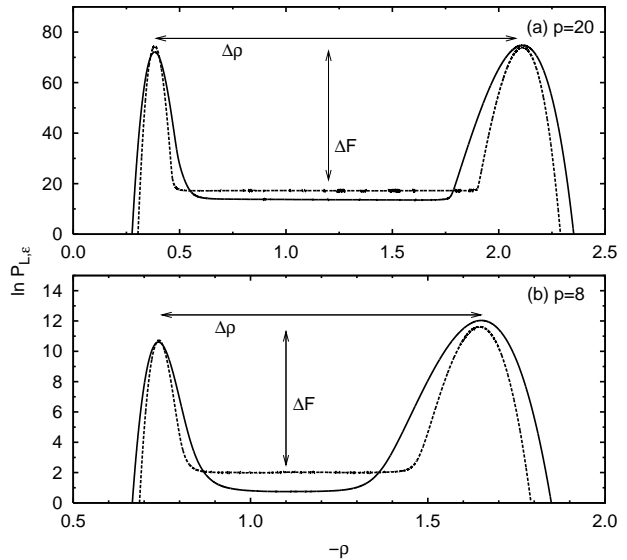} 
\caption{\label{fig:stretches} Plots of $\ln P$ as obtained in completely 
periodic simulation boxes of size $10 \times 10 \times 30$ (solid lines) and $10 
\times 10 \times 60$ (dashed lines) for (a) $p=20$ and (b) $p=8$. The height of 
the peaks $\Delta F$ is related to the interfacial tension $\gamma_\infty$ via 
\eq{eq:binder}. The distance $\Delta \rho$ between the peaks is a measure of the 
latent heat density $\cal L_\infty$. In these plots $\epsilon$ was tuned to 
yield an approximately horizontal region between the peaks.}
\end{center}
\end{figure}

Similar to previously, bulk $H \to \infty$ results are obtained using $L \times 
L \times L$ systems with periodic boundaries in all directions. The bulk latent 
heat density ${\cal L}_\infty$ is obtained from the specific heat maximum using 
\eq{eq:cvlh} and is once again extrapolated to $L \to \infty$, where now 
$N=L^3$. The resulting estimate of $\cal L_\infty$ is also listed in 
\tab{tab:first}. To obtain the bulk interfacial tension $\gamma_\infty$ we use 
the method of Binder \cite{binder:1982}. Simulating a large and stretched $L 
\times L \times D$ system $D > L$ with periodic boundaries in all directions, 
the logarithm of the energy distribution $P(E)$ reveals a pronounced flat region 
between the peaks, see \fig{fig:stretches}. The flat region indicates that the 
isotropic and nematic phase coexist with only small interactions between the two 
interfaces. Hence, the average peak height $\Delta F$ is related to the bulk 
interfacial tension
\begin{equation}\label{eq:binder}
 \gamma_\infty = \lim_{L \to \infty} \gamma_L, \hspace{5mm}
 \gamma_L = \Delta F / (2L^2),
\end{equation}
yielding an elegant method of obtaining~$\gamma_\infty$. Provided $L$ is large 
enough, the result should not depend on the elongation $D$, but inspection of 
\fig{fig:stretches} reveals this is not quite true, especially for $p=8$. 
This could indicate that some interaction between the interfaces remains, or 
that $L$ was not large enough. In any case, using the largest available system 
size, we obtain $\gamma_\infty \approx 0.05$ for $p=8$ and $\gamma_\infty 
\approx 0.29$ for $p=20$ (in units of $k_B T$ per lattice spacing squared). 
Alternatively, $\gamma_L$ can be measured in a cubic periodic system of size 
$L$ and extrapolation to $L \to \infty$ using
\begin{equation}\label{eq:gamma}
 \gamma_L = \gamma_\infty + c_1 \ln L / L^2 + c_2 / L^2,
\end{equation}
with constants $c_i$, can be attempted \cite{binder:1982}. When using this 
procedure we obtain slightly higher values of the interfacial tension, namely 
$\gamma_\infty \approx 0.08$ and $\gamma_\infty \approx 0.31$ for $p=8$ and 
$p=20$ respectively. Hence, for $p=20$ the estimates for $\gamma_\infty$ agree 
reasonably well, whereas for $p=8$ some discrepancy remains. In \tab{tab:first} 
the average of both estimates is provided.

\begin{figure}
\begin{center}
\includegraphics[width=\figwidth]{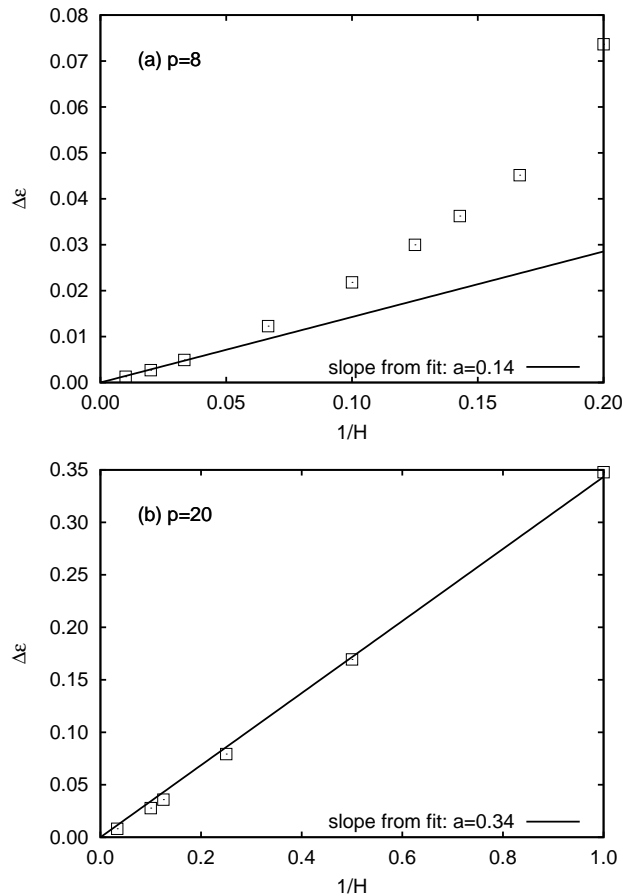}

\caption{\label{fig:kelvin} Test of the Kelvin equation. Plotted is the inverse 
temperature shift $\Delta \epsilon$ of \eq{eq:kelvin} versus the inverse film 
thickness $1/H$, using exponents $p=8$ (a) and $p=20$ (b) in \eq{eqll}.}

\end{center}
\end{figure}

We now have all quantities needed to put the Kelvin equation to the test, see 
\eq{eq:kelvin}. Shown in \fig{fig:kelvin} is $\Delta \epsilon$ versus $1/H$, for
both $p=8$ and $p=20$, using only values of $H$ where the transition is 
first-order. Provided the Kelvin equation holds, the resulting plots should be 
linear. For $p=8$ this is clearly not the case; only in the limit $1/H \to 0$, 
i.e.~where the transition is strongly first-order, is agreement observed. In 
contrast, using $p=20$ the Kelvin equation holds for all values of the film 
thickness, including $H=1$. The slope $a$ of the lines in \fig{fig:kelvin} can 
be obtained from a fit; following \eq{eq:kelvin} it is expected that $a = 2 
\gamma_\infty / \cal L_\infty$, allowing for a stringent quantitative test. For 
$p=20$ we obtain by fitting $a \approx 0.34$, which is in excellent agreement 
with $2\gamma_\infty / {\cal L}_\infty \approx 0.35$ calculated using the 
independent estimates of \tab{tab:first}. For $p=8$ the fit yields $a \approx 
0.14$, where only the largest three values of $H$ were used. Once again, this 
is in excellent agreement with $2\gamma_\infty / {\cal L}_\infty \approx 0.13$ 
obtained from \tab{tab:first}.

\section{Discussion and Summary}
\label{the_end}

In this paper we have provided new results regarding the IN transition in liquid 
crystals confined between neutral walls. The main conclusion to be taken from 
this work is that a single universal scenario describing the nature of this 
transition as function of the film thickness $H$ does not exist. Using a 
generalized version of the LL model, we have explicitly demonstrated that the 
first-order IN transition can terminate at a critical thickness $\hx$, below 
which it becomes continuous, or that it can stay first-order irrespective of 
$H$. The scenario that takes place is determined by a single parameter in the 
Hamiltonian, namely $p$ in \eq{eqll}, which sets the \ahum{sharpness} of the 
pair interaction. When the transition is sufficiently strongly first-order 
excellent agreement with the Kelvin equation is also obtained. In particular, we 
not only observe the $1/H$ shift of the transition inverse temperature but also 
the prefactor of the shift is in quantitative agreement with the {\it 
independently measured} bulk latent heat and interfacial tension. However, when 
the IN transition is only weakly first-order clear deviations appear and the 
Kelvin equation significantly underestimates the inverse temperature shift, see 
\fig{fig:kelvin}(a).

The two different manifestations of the confined IN transition presented in this 
work yield two distinct phase diagram topologies: one where the isotropic and 
nematic branches of the binodal terminate at the critical thickness $\hx$ and 
one where they continue irrespective of $H$. It is of some interest to compare 
the resulting phase diagrams to other works. The topology of the $p=8$ phase 
diagram, see \fig{phase}, is commonly encountered in confined colloidal rods and 
plates \cite{citeulike:3683414, citeulike:4922416, citeulike:3687040, 
dijkstra.roij.ea:2001}. To facilitate the comparison, the energy density in 
\fig{phase} should be interpreted as the analogue of the particle density in 
colloidal systems. In agreement with \fig{phase}, the first-order IN transition 
in colloidal systems also terminates at a critical thickness 
\cite{citeulike:3683414, citeulike:4922416, citeulike:3687040, 
dijkstra.roij.ea:2001}. It is also interesting to see that the nematic branch of 
the binodal in \fig{phase} shows rather extreme outward curvature as the bulk 
limit is approached. Colloidal platelets reveal similar behavior, albeit that 
here the effect appears in the isotropic branch \cite{citeulike:3687040}. In 
contrast with colloidal systems is the fact that \eq{eqll} with $p=8$ in the 
bulk limit yields a first-order transition that is too strong. Defining the 
relative strength of the transition as
\begin{equation}
 r = \frac{ \rho_{\rm nem} - \rho_{\rm iso} }{ 
 \rho_{\rm nem} + \rho_{\rm iso} },
\end{equation}
we obtain $r \approx 0.38$ for \eq{eqll} with $p=8$, while Onsager's exact 
solution \cite{onsager:1949} for infinitely slender rods yields $r \approx 
0.12$. This discrepancy can be fixed by using a lower $p$ in \eq{eqll}. For 
instance, $p=5$ gives $r \approx 0.15$ \cite{citeulike:5202797}, which is much 
closer to Onsager's result. Note that $p=5$ still exceeds the original LL value 
$p=2$. Indeed, it has been pointed out that the original LL model yields a bulk 
IN transition that is too weakly first-order compared to what is observed in 
fluids of rods \cite{citeulike:4484545}.

The second phase diagram topology, where the binodal branches do not terminate 
in thin films, is obtained for $p=20$ in \eq{eqll}, see \fig{fig:p20_pd}. The 
resulting phase diagram is of fundamental importance, since it clearly 
demonstrates that first-order IN transitions in thin films are also possible and 
that the crossover to a continuous transition need not necessarily take place. 
It is interesting that experiments so far have not produced clear evidence of a 
continuous IN transition in thin films \cite{citeulike:4067023, 
citeulike:4006159, citeulike:2811025, citeulike:3991276}. This is consistent 
with a phase diagram topology as shown in \fig{fig:p20_pd}. However, it is 
obvious that the model of \eq{eqll} with $p=20$ does not capture the bulk limit 
correctly, since the bulk IN transition ought to be weak, whereas $p=20$ yields 
a very strong first-order transition. Clearly, some features are still lacking 
in \eq{eqll}, for instance a coupling between the orientational and spatial 
degrees of freedom of the particles, as well as anchoring effects at the walls. 
Investigations which incorporate these effects are possible directions for 
future work.

Finally, using $p=8$ and $H<\hx$ our results show that a genuine continuous IN 
transition can also take place. Since long-range nematic order is not observed 
a transition of the KT type \cite{kosterlitz.thouless:1972} is the most likely 
scenario. This result is interesting because using $p=2$ one finds that 
\eq{eqll} is without any kind of phase transition in the thin-film limit 
\cite{citeulike:5158181, citeulike:5158109}. Hence, the nature of the IN 
transition in thin films is ultimately determined by microscopic details. This 
means that a single universality class for the IN transition cannot exist. 
Depending on the details of the interaction, there can be both first-order 
and continuous transitions as well as no transition occurring at all.

\acknowledgments

This work was supported by the {\it Deutsche Forschungsgemeinschaft}
under the Emmy Noether program (VI~483/1-1).

\bibliography{mc1975,notes}

\begin{thebibliography}{10}
\bibitem{note1}
Note that we use the inverse temperature, and also that the definition of the
  LL model in \eq{eqll} differs from the usual one by a factor of 3/2.
\end{thebibliography}

\end{document}